\documentclass{jpsj2}

\title{%
Boltzmann Equations for Spin and Charge Relaxations
in Superconductors
}

\author{%
Yositake {\sc Takane}
}

\inst{%
Department of Quantum Matter, Graduate School of Advanced Sciences of Matter,
Hiroshima University, Higashi-Hiroshima 739-8530, Japan
}

\recdate{ \hspace{50mm} }

\abst{%
In a superconductor coupled with a ferromagnetic metal,
spin and charge imbalances can be induced by injecting spin-polarized electron
current from the ferromagnetic metal.
We theoretically study a nonequilibrium distribution of quasiparticles
in the presence of spin and charge imbalances.
We show that four distribution functions are needed to characterize
such a nonequilibrium situation, and derive a set of linearized Boltzmann
equations for them by extending the argument by Schmid and Sch\"{o}n based on
the quasiclassical Green's function method.
Using the Boltzmann equations, we analyze the spin imbalance in
a thin superconducting wire weakly coupled with a ferromagnetic electrode.
The spin imbalance induces a shift $\delta\mu$ ($- \delta \mu$) of
the chemical potential for up-spin (down-spin) quasiparticles.
We discuss how $\delta \mu$ is relaxed by spin-orbit impurity scattering.
}

\kword{%
nonequilibrium superconductivity, spin imbalance, charge imbalance,
spin-orbit interaction}

\begin{document}

\maketitle

\section{Introduction}

Electron current injected into a superconductor produces a nonequilibrium
distribution of quasiparticles.~\cite{rf:rieger, rf:yu1,rf:clarke1,rf:tinkham1,
rf:tinkham2,rf:clarke2,rf:schmid,rf:yu2,rf:ikebuchi}
A number difference between electrons and holes arises
in such a nonequilibrium situation.~\cite{rf:tinkham1,rf:tinkham2}
This is called charge imbalance.
The charge imbalance induces an excess quasiparticle current,
which results in a potential difference
between pairs and quasiparticles.~\cite{rf:rieger,rf:tinkham1,rf:tinkham2}
Although early experiments on the charge imbalance have focused on
spin-independent phenomena, we can now study spin-related ones
by injecting spin-polarized electron current into superconductors
by using a ferromagnetic metal as an injection electrode.~\cite{rf:tedrow}
If injected electron current is spin polarized,
there arises spin imbalance between up- and down-spin quasiparticles
in addition to the charge imbalance.
In this case, a potential difference between up- and down-spin quasiparticles
arises in addition to a spin-independent potential difference
due to the charge imbalance.
The spin imbalance in mesoscopic superconductors has recently
attracted much attention both experimentally~\cite{rf:johnson,rf:urech,rf:shin}
and theoretically.~\cite{rf:zhao,rf:yamashita}
The main attention is focused on its relaxation due to spin-flip processes.

The most general framework to understand the charge imbalance
in superconductors is presented by Schmid and Sch\"{o}n.~\cite{rf:schmid}
Based on kinetic equations for the quasiclassical Green's functions,
they showed that nonequilibrium quasiparticle distributions are described by
two distribution functions $f_{{\rm T}}(\mib{r},\epsilon)$ and
$f_{{\rm L}}(\mib{r},\epsilon)$, where $\epsilon$ represents quasiparticle
energy measured from the chemical potential $\mu$ at equilibrium,
and derived a set of linearized Boltzmann equations for them.
It has turned out that $f_{\rm T}$ and $f_{\rm L}$ describe the charge
imbalance and the related excess quasiparticle energy, respectively,
and that they are coupled with respectively
transverse and longitudinal variations of the pair potential.
By using the Boltzmann equations, we can describe the charge imbalance
including effects of inelastic phonon scattering.
However, their framework is restricted to spin-independent phenomena,
and cannot apply to the case where the spin imbalance plays a role.
To fully understand nonequilibrium quasiparticle distributions in the presence
of a spin-polarized current injection, we need a framework by which we can
describe both the spin and charge imbalances in a unified manner
including effects of phonon scattering.
Although a few theoretical treatments for the spin imbalance in superconductors
have been reported so far,~\cite{rf:zhao,rf:yamashita}
they do not satisfy all our requirements.

In this paper, we study a nonequilibrium quasiparticle distribution
in the presence of the spin and charge imbalances.
To describe such a nonequilibrium situation,
we adopt the kinetic equation approach by Schmid and Sch\"{o}n.
We introduce four distribution functions $f_{{\rm L}+}$, $f_{{\rm L}-}$,
$f_{{\rm T}+}$ and $f_{{\rm T}-}$ to characterize quasiparticle distributions.
It is shown that $f_{{\rm L}+}$ and $f_{{\rm T}-}$ represent
the spin and charge imbalances, respectively, and $f_{{\rm L}-}$ and
$f_{{\rm T}+}$ represent the excess quasiparticle energy and the energy
imbalance between up- and down-spin quasiparticles, respectively.
The suppression of the pair potential due to a current injection
is determined by $f_{{\rm L}-}$.
We derive a set of linearized Boltzmann equations
for the distribution functions in steady states 
assuming that the spin imbalance is relaxed by spin-orbit impurity scattering.
We show that the four distribution functions are decoupled with each other
in the resulting Boltzmann equations.
This indicates that we can separately consider the spin and charge imbalances
in steady states.
As an application of the Boltzmann equations,
we treat the spin imbalance in a quasi-one-dimensional superconducting wire
weakly coupled with a ferromagnetic electrode through a tunnel junction.
The spin imbalance induces a shift $\delta\mu$ ($- \delta \mu$) of
the chemical potential for up-spin (down-spin) quasiparticles.
We discuss spatial decay of $\delta \mu$ due to spin-orbit impurity scattering.
At high temperatures, the decay of $\delta\mu(x)$ obeys the exponential law
${\rm e}^{- |x-x_{0}|/\lambda_{\rm s}}$, where $\lambda_{\rm s}$ is
the spin-diffusion length and $|x-x_{0}|$ represents the distance from
the injection point $x_{0}$.
This is in agreement with the previous result.~\cite{rf:yamashita}
However, at low temperatures,
we observe deviations from the exponential law near the injection point.

In the next section, 
we consider a superconductor coupled with a ferromagnetic metal and
introduce the quasiclassical Green's functions for the superconductor
in the Keldysh formalism.
The kinetic equations for them are derived in the presence of
spin-orbit impurity scattering.
In \S 3, we introduce four distribution functions to describe both the spin
and charge imbalances, and derive a set of Boltzmann equations.
We clarify the meaning of each distribution function.
In \S 4, we analyze the spin-imbalance relaxation in a thin superconducting
wire based on the resulting Boltzmann equations.
Section 5 is devoted to a short summary.
We set $\hbar = k_{\rm B} = 1$ throughout this paper.

\section{Kinetic Equations for Green's Functions}

We consider a superconductor coupled with a ferromagnetic metal
through a point-like tunnel junction.
As a model for the superconductor, we adopt an electron gas
interacting with phonons.
Phonons are assumed to be described by the Debye model
with the sound velocity $c_{0}$.
We assume that electrons experience normal impurity scattering
and spin-orbit impurity scattering.
Let $\psi_{\sigma}(x)$ with $x \equiv (\mib{r},t)$
be the electron field operator in the superconductor.
We introduce $\Psi(x) = \hspace{1mm}^{t}( \psi_{\uparrow}(x),
\psi_{\downarrow}^{\dagger}(x) )$ and
define the following Green's functions~\cite{rf:kopnin}
\begin{align}
    \label{eq:def-G1}
  \hat{G}^{K}(x,x') & = - {\rm i} \hat{\tau}_{z}
     \big\langle \big[ \Psi(x), \Psi^{\dagger}(x') \big]_{-}
     \big\rangle ,
          \\
    \label{eq:def-G2}
  \hat{G}^{R}(x,x') & = - {\rm i} \hat{\tau}_{z} \Theta(t-t')
     \big\langle \big[ \Psi(x), \Psi^{\dagger}(x') \big]_{+}
     \big\rangle ,
          \\
    \label{eq:def-G3}
  \hat{G}^{A}(x,x') & = + {\rm i} \hat{\tau}_{z} \Theta(t'-t)
     \big\langle \big[ \Psi(x), \Psi^{\dagger}(x') \big]_{+}
     \big\rangle ,
\end{align}
where $\Theta(t)$ is Heavisid's step function.
Here and hereafter, $\hat{\tau_{i}}$ ($i = x,y,z$) represents the Pauli matrix.
We use the Keldysh representation
\begin{align}
    \label{eq:keldysh-rp}
 \underline{G}(x,x')
  = \left( \begin{array}{cc}
              \hat{G}^{R}(x,x') & \hat{G}^{K}(x,x') \\
                      0         & \hat{G}^{A}(x,x')
           \end{array}
      \right) ,
\end{align}
and define its Fourier transform as
\begin{align}
  \underline{G}(\mib{r},\mib{p},t,t') \equiv
   \int {\rm d}^{3}s \hspace{1mm} {\rm e}^{-{\rm i}p s}
   \underline{G} \left(\mib{r}+\frac{\mib{s}}{2},t,
                        \mib{r}-\frac{\mib{s}}{2},t'\right) .
\end{align}
Integrating this over $\xi \equiv \mib{p}^{2}/(2m)-\mu$,
we obtain the quasiclassical Green's function~\cite{rf:eilenberger}
\begin{align}
 \underline{G}(\mib{r},\hat{\mib{p}},t,t') =
   \frac{{\rm i}}{\pi} \int {\rm d}\xi \hspace{1mm}
   \underline{G}(\mib{r},\mib{p},t,t') ,
\end{align}
where $\hat{\mib{p}} = \mib{p}/|\mib{p}|$.
It has been shown that the quasiclassical Green's function
obeys~\cite{rf:eliashberg,rf:larkin}
\begin{align}
       \label{eq:qc-equation1}
&  v_{\rm F} \hat{\mib{p}} \cdot \nabla
                \underline{G}(\mib{r},\hat{\mib{p}},t,t')
 + \underline{\tau}_{z} \partial_{t} \underline{G}(\mib{r},\hat{\mib{p}},t,t')
 + \partial_{t'} \underline{G}(\mib{r},\hat{\mib{p}},t,t') \underline{\tau}_{z}
                 \nonumber \\
& \hspace{10mm}
 + {\rm i} \int_{-\infty}^{\infty} {\rm d} t_{1}
           \left(  \underline{\Sigma}(\mib{r},\hat{\mib{p}},t,t_{1})
                   \underline{G}(\mib{r},\hat{\mib{p}},t_{1},t')
                 - \underline{G}(\mib{r},\hat{\mib{p}},t,t_{1})
                   \underline{\Sigma}(\mib{r},\hat{\mib{p}},t_{1},t')
           \right)
 = 0 ,
\end{align}
where
$\underline{\tau}_{i} \equiv {\rm diag}(\hat{\tau}_{i}, \hat{\tau}_{i})$
and $\underline{\Sigma}(\mib{r},\hat{\mib{p}},t,t')$ represents
the self-energy part.
The self-energy part is decomposed into
\begin{align}
        \label{eq:self-energy1}
    \underline{\Sigma}(\mib{r},\hat{\mib{p}},t,t')
 = \underline{\Sigma}_{\rm imp}(\mib{r},\hat{\mib{p}},t,t')
     + \underline{\Sigma}_{\rm so}(\mib{r},\hat{\mib{p}},t,t')
     + \underline{\Sigma}_{\rm ph}(\mib{r},\hat{\mib{p}},t,t')
     + \underline{\Sigma}_{\rm inj}(\mib{r},\hat{\mib{p}},t,t') ,
\end{align}
where $\underline{\Sigma}_{\rm imp}$, $\underline{\Sigma}_{\rm so}$ and
$\underline{\Sigma}_{\rm ph}$ represent the contributions from
normal impurity scattering, spin-orbit impurity scattering
and electron-phonon interaction, respectively.
It should be noted that $\underline{\Sigma}_{\rm ph}$ contains
the pair potential.
The last term $\underline{\Sigma}_{\rm inj}$ represents the spin-polarized
current injection from the ferromagnetic metal.
We simplify eq.~(\ref{eq:qc-equation1}) by following
the argument by Usadel.~\cite{rf:usadel}
We employ an approximation
\begin{align}
    \underline{G}(\mib{r},\hat{\mib{p}},t,t')
  = \underline{G}(\mib{r},t,t')
         + \hat{\mib{p}} \cdot \underline{G}_{1}(\mib{r},t,t') ,
\end{align}
where $\underline{G}_{1}(\mib{r},t,t')$ is much smaller then
$\underline{G}(\mib{r},t,t')$.
After some manipulations, we obtain
\begin{align}
     \label{eq:qc-equation2}
&   D \nabla \cdot \left( \underline{G}(\mib{r},\epsilon,t)
                          \nabla \underline{G}(\mib{r},\epsilon,t) \right)
  + {\rm i}\epsilon \left[ \underline{\tau}_{z},
                           \underline{G}(\mib{r},\epsilon,t) \right]_{-}
         \nonumber \\
& \hspace{10mm}
  - \frac{1}{2} \partial_{t}
    \left[ \underline{\tau}_{z}, \underline{G}(\mib{r},\epsilon,t) \right]_{+}
  - {\rm i} \left[ \underline{\Sigma}(\mib{r},\epsilon,t),
                   \underline{G}(\mib{r},\epsilon,t) \right]_{-}
  = 0 ,
\end{align}
where
\begin{align}
     \underline{G}(\mib{r},\epsilon,t)
   = \int {\rm d} s \hspace{1mm} {\rm e}^{{\rm i}\epsilon s}
     \underline{G}\left(\mib{r},t+\frac{s}{2},t-\frac{s}{2}\right) .
\end{align}
In the following, we restrict our consideration to steady states,
and neglect the $t$-dependence of all the functions.

We evaluate the elements of the self-energy part.
We first consider $\underline{\Sigma}_{\rm imp}$ and
$\underline{\Sigma}_{\rm so}$.
The Hamiltonian $H_{\rm imp}$ for normal impurity scattering is written as
\begin{align}
  H_{\rm imp}
    = \sum_{\sigma} \int {\rm d}^{3}r \hspace{1mm}
      \psi_{\sigma}^{\dagger}(\mib{r})V_{\rm imp}(\mib{r})
      \psi_{\sigma}(\mib{r}) .
\end{align}
We assume for simplicity that the impurity potential is given by
\begin{align}
           \label{eq:imp-pot}
  V_{\rm imp}(\mib{r}) = u_{\rm imp} \sum_{j} \delta (\mib{r}-\mib{R}_{j}) ,
\end{align}
where $\mib{R}_{j}$ indicates the position of the $j$th impurity.
The Hamiltonian $H_{\rm so}$ for spin-orbit impurity scattering is written as
\begin{align}
  H_{\rm so}
    = \sum_{\sigma, \sigma'} \int {\rm d}^{3}r \hspace{1mm}
      \psi_{\sigma}^{\dagger}(\mib{r})
      U_{\rm so}^{\sigma,\sigma'}(\mib{r})
      \psi_{\sigma'}(\mib{r}) .
\end{align}
The spin-dependent potential is
\begin{align}
  U_{\rm so}^{\sigma,\sigma'}(\mib{r})
     = - {\rm i} (\vec{\hat{\tau}})_{\sigma,\sigma'} \cdot
       \left( \nabla V_{\rm so}(\mib{r}) \times \nabla \right) ,
\end{align}
where $V_{\rm so}(\mib{r})$ is given by eq.~(\ref{eq:imp-pot}) with
the replacement of $u_{\rm imp} \to u_{\rm so} / k_{\rm F}^{2}$.
In this case, $\underline{\Sigma}^{\rm imp}$ and $\underline{\Sigma}^{\rm so}$
are given by
\begin{align}
  \underline{\Sigma}_{\rm imp}(\mib{r},\epsilon)
 & = - \frac{{\rm i}}{2\tau_{\rm imp}} \underline{G}(\mib{r},\epsilon) ,
            \\
  \underline{\Sigma}_{\rm so}(\mib{r},\epsilon)
 & = - \frac{{\rm i}}{6\tau_{\rm so}} \underline{G}(\mib{r},\epsilon)
     + \frac{{\rm i}}{3\tau_{\rm so}} \underline{\tau}_{y}
       \underline{G}^{'}(\mib{r},\epsilon) \underline{\tau}_{y} .
\end{align}
The relaxation times are given by
\begin{align}
     \frac{1}{\tau_{\rm imp}}
 & = 2\pi n_{\rm imp}N_{\rm S}(0)|u_{\rm imp}|^{2} ,
           \\
     \frac{1}{\tau_{\rm so}}
 & = 2\pi n_{\rm imp}N_{\rm S}(0)|u_{\rm so}|^{2}\sum_{i = x,y,z}
     \overline{(\hat{\mib{p}}\times\hat{\mib{p}}^{'})_{i}^{2}} ,
\end{align}
where $n_{\rm imp}$ and $N_{\rm S}(0)$ represent the impurity concentration
and the density of states in the normal state, respectively.
Note that the second term of $\underline{\Sigma}^{\rm so}$ with
$\underline{G}^{'}$ represents spin-flip processes.
The Green's function $\underline{G}^{'}(\mib{r},\epsilon)$
is the Fourier transform of $\underline{G}^{'}(x,x')$,
where $\underline{G}^{'}(x,x')$ is defined by
eqs.~(\ref{eq:def-G1})-(\ref{eq:keldysh-rp})
with the replacement $\Psi(x) \to \Psi^{'}(x)
= \hspace{1mm}^{t}( \psi_{\uparrow}^{\dagger}(x), \psi_{\downarrow}(x) )$.
We can show that its elements $\hat{G}^{'X}(x,x')$ ($X = R, A, K$) satisfy
\begin{align}
       \label{eq:G'-G}
  \hat{G}^{'X}(x,x')
 = - \hat{\tau}_{z} \hspace{1mm}
     ^{t} [\hat{G}^{\bar{X}}(x',x)] \hat{\tau}_{z} ,
\end{align}
where $\bar{R} = A$, $\bar{A} = R$ and $\bar{K} = K$.
These relation will be used later.
We obtain
\begin{align}
     \underline{\Sigma}_{\rm imp}(\mib{r},\epsilon)
   + \underline{\Sigma}_{\rm so}(\mib{r},\epsilon)
  = - \frac{{\rm i}}{2\tau} \underline{G}(\mib{r},\epsilon)
    - \frac{{\rm i}}{3\tau_{\rm so}} 
      \left( - \underline{\tau}_{y} \underline{G}^{'}(\mib{r},\epsilon)
               \underline{\tau}_{y} - \underline{G}(\mib{r},\epsilon)
      \right)
\end{align}
with $\tau^{-1} \equiv \tau_{\rm imp}^{-1} + \tau_{\rm so}^{-1}$.

We turn to $\underline{\Sigma}_{\rm inj}$.
Let us consider the case where a ferromagnetic metal is coupled to
the superconductor through a point-like tunnel junction at $\mib{r}_{0}$.
The corresponding Hamiltonian $H_{\rm inj}$ is given by
\begin{align}
  H_{\rm inj}
    = \sum_{\sigma} \int {\rm d}^{3}r {\rm d}^{3}r' \hspace{1mm}
      \left(  \mathcal{T}(\mib{r},\mib{r}')\psi_{\sigma}^{\dagger}(\mib{r})
              \phi_{\sigma}(\mib{r}') {\rm e}^{-{\rm i}eVt}
            + {\rm h.c.} \right) ,
\end{align}
where $\phi_{\sigma}$ and $V$ represent the electron field operator in the
ferromagnetic metal and an applied bias voltage, respectively,
and we assume that $\mathcal{T}(\mib{r},\mib{r}')
= \mathcal{T} \delta(\mib{r}-\mib{r}_{0})\delta(\mib{r}'-{\mib{r}'}_{0})$.
With this model, we obtain
\begin{align}
  \hat{\Sigma}_{\rm inj}^{X}(\mib{r},\epsilon)
        = - {\rm i}|\mathcal{T}|^{2}\delta(\mib{r}-\mib{r}_{0})
            \hat{\sigma}_{\rm inj}^{X}(\epsilon)
\end{align}
with
\begin{align}
  \hat{\sigma}_{\rm inj}^{R,A}(\epsilon)
 & =   \pm \left( \begin{array}{cc}
                    N_{{\rm F}\uparrow}(0) & 0 \\
                    0 & -N_{{\rm F}\downarrow}(0)
                  \end{array}
           \right) ,
            \\
  \hat{\sigma}_{\rm inj}^{K}(\epsilon)
 & = 2 \left( \begin{array}{cc}
                \tanh\bigl(\frac{\epsilon-eV}{2T}\bigr)
                N_{{\rm F}\uparrow}(0) & 0 \\
                0 & - \tanh\bigl(\frac{\epsilon+eV}{2T}\bigr)
                      N_{{\rm F}\downarrow}(0)
              \end{array}
       \right) ,
\end{align}
where $N_{{\rm F}\uparrow}(0)$ ($N_{{\rm F}\downarrow}(0)$) represents
the density of states for up-spin (down-spin) electrons
in the ferromagnetic metal.

The phonon self-energy part has been presented in ref~\citen{rf:eliashberg}.
Note that $\hat{\Sigma}_{\rm ph}^{R}$ and $\hat{\Sigma}_{\rm ph}^{A}$
contain the pair potential, so we separate it out as
\begin{align}
   \hat{\Sigma}_{\rm ph}^{R,A}(\mib{r},\epsilon)
 & \to \hat{\Sigma}_{\rm ph}^{R,A}(\mib{r},\epsilon)
       - {\rm i} \hat{\tau}_{\chi}(\mib{r}) \Delta(\mib{r})
\end{align}
with
\begin{align}
\hat{\tau}_{\chi}(\mib{r}) \equiv \cos \chi (\mib{r}) \hat{\tau}_{y}
+ \sin \chi (\mib{r}) \hat{\tau}_{x} ,
\end{align}
where $\chi(\mib{r})$ and $\Delta(\mib{r})$ represent the phase and amplitude
of the pair potential, respectively.
Assuming that phonons are in thermal equilibrium,
we obtain
\begin{align}
    \hat{\Sigma}_{\rm ph}^{R,A}(\mib{r},\epsilon)
 & = - {\rm i} \int{\rm d} \epsilon'\sigma_{\rm ph}(\epsilon, \epsilon')
     \Bigl\{  \coth\Bigl(\frac{\epsilon'-\epsilon}{2T}\Bigr)
              \hat{G}^{R,A}(\mib{r},\epsilon')
              \mp \frac{1}{2}
                 \hat{G}^{K}(\mib{r},\epsilon') \Bigr\} ,
                   \\
    \hat{\Sigma}_{\rm ph}^{K}(\mib{r},\epsilon)
 & = - {\rm i} \int{\rm d} \epsilon'\sigma_{\rm ph}(\epsilon, \epsilon')
     \Bigl\{  \coth\Bigl(\frac{\epsilon'-\epsilon}{2T}\Bigr)
              \hat{G}^{K}(\mib{r},\epsilon')
              - \bigl( \hat{G}^{R}(\mib{r},\epsilon')
                         - \hat{G}^{A}(\mib{r},\epsilon')
                \bigr) \Bigr\} ,
\end{align}
where
\begin{align}
      \label{eq:e-ph-coupling}
  \sigma_{\rm ph}(\epsilon, \epsilon')
  = \frac{N_{\rm S}(0)g^{2}}{8} \cdot \frac{\pi }{(c_{0}k_{\rm F})^{2}}
    (\epsilon' - \epsilon)^{2}{\rm sign}(\epsilon' - \epsilon) .
\end{align}
Here, $c_{0}$ and $g$ represent the sound velocity and the coupling constant
for the electron-phonon interaction, respectively.

Using eq.~(\ref{eq:qc-equation2}) and the elements of the self-energy part,
we obtain the kinetic equations for the Green's functions,
\begin{align}
     \label{eq:qc-equation3-1}
&   D \nabla \cdot \left( \hat{G}^{R,A} \nabla \hat{G}^{R,A} \right)
  + {\rm i}\epsilon \bigl[ \hat{\tau}_{z}, \hat{G}^{R,A} \bigr]_{-}
  - \Delta \bigl[ \hat{\tau}_{\chi}, \hat{G}^{R,A} \bigr]_{-}
  + \hat{\Gamma}^{R,A} - \hat{I}^{R,A} - \hat{P}^{R,A} = 0,
             \\
     \label{eq:qc-equation3-2}
&   D \nabla \cdot
    \bigl( \hat{G}^{R} \nabla \hat{G}^{K} + \hat{G}^{K} \nabla \hat{G}^{A}
    \bigr)
  + {\rm i}\epsilon \bigl[ \hat{\tau}_{z}, \hat{G}^{K} \bigr]_{-}
  - \Delta \bigl[ \hat{\tau}_{\chi}, \hat{G}^{K} \bigr]_{-}
  + \hat{\Gamma}^{K} - \hat{I}^{K} - \hat{P}^{K} = 0,
\end{align}
where
\begin{align}
          \label{eq:Gamma-RA}
      \hat{\Gamma}^{R,A}
  & = \frac{1}{3\tau_{\rm so}} 
      \bigl( \hat{\tau}_{y} \hat{G}^{'R,A} \hat{\tau}_{y}\hat{G}^{R,A}
             - \hat{G}^{R,A} \hat{\tau}_{y} \hat{G}^{'R,A} \hat{\tau}_{y}
      \bigr) ,
         \\
          \label{eq:Gamma-K}
      \hat{\Gamma}^{K}
  & = \frac{1}{3\tau_{\rm so}} 
      \bigl( \hat{\tau}_{y} \hat{G}^{'R} \hat{\tau}_{y} \hat{G}^{K}
           + \hat{\tau}_{y} \hat{G}^{'K} \hat{\tau}_{y} \hat{G}^{A}
           - \hat{G}^{R} \hat{\tau}_{y} \hat{G}^{'K} \hat{\tau}_{y}
           - \hat{G}^{K} \hat{\tau}_{y} \hat{G}^{'A} \hat{\tau}_{y}
      \bigr) ,
        \\
     \hat{I}^{R, A}
  & = {\rm i} \bigl( \hat{\Sigma}_{\rm ph}^{R,A} \hat{G}^{R,A}
                     - \hat{G}^{R,A} \hat{\Sigma}_{\rm ph}^{R,A} \bigr) ,
         \\
      \hat{I}^{K}
  & = {\rm i} \bigl(  \hat{\Sigma}_{\rm ph}^{R} \hat{G}^{K}
                    + \hat{\Sigma}_{\rm ph}^{K} \hat{G}^{A}
                    - \hat{G}^{R} \hat{\Sigma}_{\rm ph}^{K}
                    - \hat{G}^{K} \hat{\Sigma}_{\rm ph}^{A}
              \bigr) ,
\end{align}
and $\hat{P}^{X}$ is obtained by the replacement of
$\hat{\Sigma}_{\rm ph}^{X} \to \hat{\Sigma}_{\rm inj}^{X}$
in $\hat{I}^{X}$.
The arguments $\mib{r}$ and $\epsilon$ are suppressed in the above equations.
Here, $\hat{\Gamma}^{X}$ represent the influence of spin-flip processes
due to spin-orbit impurity scattering, and $\hat{I}^{X}$ and $\hat{P}^{X}$
describe inelastic phonon scattering and a spin-polarized current injection,
respectively.
If we neglect $\hat{\Gamma}^{X}$ and set
$N_{{\rm F}\uparrow}(0) = N_{{\rm F}\downarrow}(0)$ in $\hat{P}^{X}$,
our argument is reduced to
the previous one presented by Schmid and Sch{\"o}n.~\cite{rf:schmid}
Quasiparticle distribution functions are contained in $\hat{G}^{K}$
and $\hat{G}^{'K}$.

\section{Boltzmann Equations}

Based on the kinetic equations presented in the previous section,
we start to derive a set of linearized Boltzmann equations
which describes a nonequilibrium quasiparticle distribution in superconductors.
In terms of the spectral functions $N_{1}$, $N_{2}$, $R_{1}$ and $R_{2}$,
we approximately express $\hat{G}^{R}$ and $\hat{G}^{A}$ as~\cite{rf:schmid} 
\begin{align}
        \label{eq:G-NR}
  \hat{G}^{R,A}(\mib{r},\epsilon)
 =   \bigl(\pm N_{1}(\mib{r},\epsilon) + {\rm i} R_{1}(\mib{r},\epsilon) \bigr)
     \hat{\tau}_{z}
   + \bigl(N_{2}(\mib{r},\epsilon) \pm {\rm i} R_{2}(\mib{r},\epsilon) \bigr)
     \hat{\tau}_{\chi}(\mib{r}) .
\end{align}
The spectral functions satisfy
\begin{align}
        \label{eq:N-even}
   N_{1,2}(\mib{r},\epsilon)
 & = N_{1,2}(\mib{r},-\epsilon) ,
         \\
        \label{eq:R-odd}
   R_{1,2}(\mib{r},\epsilon)
 & = - R_{1,2}(\mib{r},-\epsilon) .
\end{align}
We rewrite $\hat{\tau}_{y}\hat{G}^{'X}(\mib{r}, \epsilon)\hat{\tau}_{y}$
in terms of $\hat{G}^{X}(\mib{r}, \epsilon)$.
From eq.~(\ref{eq:G'-G}), we obtain
\begin{align}
    \label{eq:G'-G2}
       \hat{G}^{'X}(\mib{r}, \epsilon)
   = - \hat{\tau}_{z} \hspace{1mm}
       ^{t}[\hat{G}^{\bar{X}}(\mib{r}, - \epsilon)]
       \hspace{1mm} \hat{\tau}_{z} .
\end{align}
Combining eqs.~(\ref{eq:G-NR}) and (\ref{eq:G'-G2})
with eqs.~(\ref{eq:N-even}) and (\ref{eq:R-odd}), we obtain
$\hat{\tau}_{y}\hat{G}^{'R}(\mib{r}, \epsilon)\hat{\tau}_{y}
= - \hat{G}^{R}(\mib{r}, \epsilon)$ and
$\hat{\tau}_{y}\hat{G}^{'A}(\mib{r}, \epsilon)\hat{\tau}_{y}
= - \hat{G}^{A}(\mib{r}, \epsilon)$.
Note that $\hat{\Gamma}^{R,A} = 0$ if these results are substituted
into eq.~(\ref{eq:Gamma-RA}).
This indicates that spin-flip processes due to spin-orbit impurity scattering
do not influence on the spectral properties.~\cite{rf:abrikosov}
Only quasiparticle distribution functions contained in $\hat{G}^{K}$
are affected by spin-flip processes.
We adopt the following expression~\cite{rf:larkin}
\begin{align}
      \label{eq:G-K}
   \hat{G}^{K}(\mib{r},\epsilon)
 = \hat{G}^{R}(\mib{r},\epsilon) \hat{h}(\mib{r},\epsilon)
        - \hat{h}(\mib{r},\epsilon) \hat{G}^{A}(\mib{r},\epsilon)
\end{align}
with
\begin{align}
    \hat{h}(\mib{r},\epsilon)
  = h_{1}(\mib{r},\epsilon) + h_{2}(\mib{r},\epsilon)\hat{\tau}_{z} .
\end{align}
A nonequilibrium quasiparticle distribution is described by
$h_{1}$ and $h_{2}$.
From eq.~(\ref{eq:G'-G2}), we obtain
\begin{align}
       \label{eq:G'-K}
   \hat{\tau}_{y}\hat{G}^{'K}(\mib{r}, \epsilon)\hat{\tau}_{y}
   =  \hat{G}^{R}(\mib{r}, \epsilon)
      \hat{\tau}_{y} \hat{h}(\mib{r}, - \epsilon) \hat{\tau}_{y}
   -  \hat{\tau}_{y} \hat{h}(\mib{r}, - \epsilon) \hat{\tau}_{y}
      \hat{G}^{A}(\mib{r}, \epsilon) .
\end{align}
By using eqs.~(\ref{eq:G-K}) and (\ref{eq:G'-K}),
we simplify the expression of $\hat{\Gamma}^{K}$ as
\begin{align}
       \label{eq:gaama-K}
      \hat{\Gamma}^{K}(\mib{r},\epsilon)
    = \frac{2}{3\tau_{\rm so}} 
      \left( \hat{G}^{R}(\mib{r},\epsilon) \delta \hat{h}(\mib{r},\epsilon)
             \hat{G}^{A}(\mib{r},\epsilon) - \delta \hat{h}(\mib{r},\epsilon)
      \right)
\end{align}
with
\begin{align}
      \delta \hat{h}(\mib{r},\epsilon)
    =   \left( h_{1}(\mib{r},\epsilon)+h_{1}(\mib{r},- \epsilon) \right)
      + \left( h_{2}(\mib{r},\epsilon)-h_{2}(\mib{r},- \epsilon) \right)
        \hat{\tau}_{z} .
\end{align}
At equilibrium, $h_{1,2}(\mib{r},\epsilon)$ is reduced to
$h_{1} = \tanh (\epsilon/(2T))$ and $h_{2} = 0$.
It is convenient to set
\begin{align}
 h_{1}(\mib{r},\epsilon) & = \tanh \Bigl(\frac{\epsilon}{2T}\Bigr)
                                - 2 f_{\rm L}(\mib{r},\epsilon) ,
        \\
 h_{2}(\mib{r},\epsilon) & = - 2 f_{\rm T}(\mib{r},\epsilon) .
\end{align}
The quasiparticle distribution functions are written as~\cite{rf:schmid}
\begin{align}
   f_{\uparrow}(\mib{r},\epsilon) & = f_{\rm FD}(\epsilon,\mu)
         + f_{\rm L}(\mib{r},\epsilon) + f_{\rm T}(\mib{r},\epsilon),
            \\
   f_{\downarrow}(\mib{r},\epsilon) & = f_{\rm FD}(\epsilon,\mu)
         - f_{\rm L}(\mib{r},-\epsilon) + f_{\rm T}(\mib{r},-\epsilon) ,
\end{align}
where $f_{\rm FD}(\epsilon,\mu)$ is the Fermi-Dirac distribution function.
It should be emphasized that
$f_{\rm L}(\mib{r},\epsilon) = - f_{\rm L}(\mib{r},-\epsilon)$
and $f_{\rm T}(\mib{r},\epsilon) = f_{\rm T}(\mib{r},-\epsilon)$ are
implicitly assumed in ref.~\citen{rf:schmid}.
These relations straightforwardly result in $f_{\uparrow}(\mib{r},\epsilon)
= f_{\downarrow}(\mib{r},\epsilon)$.
Thus, the framework by Schmid and Sch{\"o}n is restricted to
spin-independent phenomena.
We do not accept the symmetry relations to enable us to consider
the spin imbalance.

We obtain Boltzmann equations based on eqs.~(\ref{eq:qc-equation3-1})
and (\ref{eq:gaama-K}).
Substituting eqs.~(\ref{eq:G-K}) and (\ref{eq:G'-K}) into
eq.~(\ref{eq:qc-equation3-2}), we derive equations
for $f_{\rm L}(\mib{r},\epsilon)$ and $f_{\rm T}(\mib{r},\epsilon)$
with the help of eq.~(\ref{eq:qc-equation3-1}).
Details of the derivation are described in ref.~\citen{rf:kopnin}.
We obtain
\begin{align}
  &  D \cdot \nabla \bigl(  N_{1}^{2}(\mib{r},\epsilon)
                          - R_{2}^{2}(\mib{r},\epsilon) \bigr)
       \nabla f_{\rm L}(\mib{r},\epsilon)
   - \frac{2}{3\tau_{\rm so}} \bigl(  N_{1}^{2}(\mib{r},\epsilon)
                                    - R_{2}^{2}(\mib{r},\epsilon) \bigr)
              \bigl(  f_{\rm L}(\mib{r},\epsilon)
                                   + f_{\rm L}(\mib{r},-\epsilon)\bigr)
                 \nonumber \\
 & \hspace{10mm}
   + I_{\rm L} \bigl(\mib{r},\epsilon,\{f_{\rm L}\}\bigr)
   + P_{\rm L} (\mib{r},\epsilon) = 0 ,
               \\
  &  D \cdot \nabla \bigl(  N_{1}^{2}(\mib{r},\epsilon)
                          + N_{2}^{2}(\mib{r},\epsilon) \bigr)
       \nabla f_{\rm T}(\mib{r},\epsilon)
   - \frac{2}{3\tau_{\rm so}} \bigl(  N_{1}^{2}(\mib{r},\epsilon)
                                    + N_{2}^{2}(\mib{r},\epsilon) \bigr)
              \bigl(  f_{\rm T}(\mib{r},\epsilon)
                                   - f_{\rm T}(\mib{r},-\epsilon)\bigr)
                 \nonumber \\
 & \hspace{10mm}
   - \frac{1}{\tau_{\rm conv}(\epsilon)} f_{\rm T}(\mib{r},\epsilon)
   + I_{\rm T} \bigl(\mib{r},\epsilon,\{f_{\rm T}\}\bigr)
   + P_{\rm T} (\mib{r},\epsilon) = 0 ,
\end{align}
where $\tau_{\rm conv}(\epsilon)$ represents
the conversion time for charge imbalance.~\cite{rf:takane}
The collision integrals $I_{\rm L}$ and $I_{\rm T}$ due to inelastic phonon
scattering are expressed as
\begin{align}
  I_{\rm L,T}\bigl(\mib{r},\epsilon,\{f\}\bigr)
 & = - 2 \int {\rm d} \epsilon' \sigma_{\rm ph}(\epsilon,\epsilon')
         M_{\rm L,T}(\mib{r},\epsilon,\epsilon')
             \nonumber \\
 & \hspace{5mm} \times
     \frac{\cosh^{2}\bigl(\frac{\epsilon}{2T}\bigr)f(\mib{r},\epsilon)
           - \cosh^{2}\bigl(\frac{\epsilon'}{2T}\bigr)
                                             f(\mib{r},\epsilon')}
          {\sinh\bigl(\frac{\epsilon'-\epsilon}{2T}\bigr)
           \cosh\bigl(\frac{\epsilon}{2T}\bigr)
           \cosh\bigl(\frac{\epsilon'}{2T}\bigr)} ,
\end{align}
where
\begin{align}
     M_{\rm L}(\mib{r},\epsilon,\epsilon')
 & = N_{1}(\mib{r},\epsilon)N_{1}(\mib{r},\epsilon')
           - R_{2}(\mib{r},\epsilon)R_{2}(\mib{r},\epsilon') ,
            \\
    M_{\rm T}(\mib{r},\epsilon,\epsilon')
 & = N_{1}(\mib{r},\epsilon)N_{1}(\mib{r},\epsilon')
           + N_{2}(\mib{r},\epsilon)N_{2}(\mib{r},\epsilon') .
\end{align}
The injection terms $P_{\rm L}$ and $P_{\rm T}$ are given by
\begin{align}
   P_{\rm L}(\mib{r},\epsilon)
& = \frac{\pi}{2}|\mathcal{T}|^{2}\delta(\mib{r}-\mib{r}_{0})
    N_{1}(\mib{r},\epsilon)
    \Bigl\{ N_{{\rm F}\uparrow}(0)
              \Bigl( \tanh\bigl(\frac{\epsilon}{2T}\bigr)
                    - \tanh\bigl(\frac{\epsilon-eV}{2T}\bigr) \Bigr)
            \nonumber \\
& \hspace{45mm}
          + N_{{\rm F}\downarrow}(0)
              \Bigl( \tanh\bigl(\frac{\epsilon}{2T}\bigr)
                   - \tanh\bigl(\frac{\epsilon+eV}{2T}\bigr) \Bigr) \Bigr\} ,
        \\
   P_{\rm T}(\mib{r},\epsilon)
& = \frac{\pi}{2}|\mathcal{T}|^{2}\delta(\mib{r}-\mib{r}_{0})
    N_{1}(\mib{r},\epsilon)
    \Bigl\{ N_{{\rm F}\uparrow}(0)
              \Bigl( \tanh\bigl(\frac{\epsilon}{2T}\bigr)
                   - \tanh\bigl(\frac{\epsilon-eV}{2T}\bigr) \Bigr)
            \nonumber \\
& \hspace{45mm}
         - N_{{\rm F}\downarrow}(0)
              \Bigl( \tanh\bigl(\frac{\epsilon}{2T}\bigr)
                   - \tanh\bigl(\frac{\epsilon+eV}{2T}\bigr) \Bigr) \Bigr\} .
\end{align}

To make the equations much simpler, we introduce the four distribution
functions,
\begin{align}
     f_{{\rm L}+}(\mib{r},\epsilon)
 & = \frac{1}{2} \left(   f_{\rm L}(\mib{r},\epsilon)
                         + f_{\rm L}(\mib{r},-\epsilon) \right) ,
       \\
     f_{{\rm L}-}(\mib{r},\epsilon)
 & = \frac{1}{2} \left(   f_{\rm L}(\mib{r},\epsilon)
                         - f_{\rm L}(\mib{r},-\epsilon) \right) ,
       \\
     f_{{\rm T}+}(\mib{r},\epsilon)
 & = \frac{1}{2} \left(   f_{\rm T}(\mib{r},\epsilon)
                         + f_{\rm T}(\mib{r},-\epsilon) \right) ,
       \\
     f_{{\rm T}-}(\mib{r},\epsilon)
 & = \frac{1}{2} \left(   f_{\rm T}(\mib{r},\epsilon)
                         - f_{\rm T}(\mib{r},-\epsilon) \right) .
\end{align}
We observe that they satisfy
\begin{align}
   f_{{\rm L,T}+}(\mib{r},-\epsilon) & = f_{{\rm L,T}+}(\mib{r},\epsilon) ,
           \\
   f_{{\rm L,T}-}(\mib{r},-\epsilon) & = - f_{{\rm L,T}-}(\mib{r},\epsilon) .
\end{align}
Noting eqs.~(\ref{eq:e-ph-coupling}), (\ref{eq:N-even}) and (\ref{eq:R-odd}),
we obtain a set of Boltzmann equations for
$f_{{\rm L}\pm}$ and $f_{{\rm T}\pm}$ as
\begin{align}
        \label{eq:fL+}
  &  D \cdot \nabla \bigl(  N_{1}^{2}(\mib{r},\epsilon)
                          - R_{2}^{2}(\mib{r},\epsilon) \bigr)
       \nabla f_{{\rm L}+}(\mib{r},\epsilon)
                 \nonumber \\
 & \hspace{10mm}
   - \frac{1}{\tau_{\rm sf}} \bigl(   N_{1}^{2}(\mib{r},\epsilon)
                                    - R_{2}^{2}(\mib{r},\epsilon) \bigr)
                             f_{{\rm L}+}(\mib{r},\epsilon)
   + I_{\rm L}\bigl(\mib{r},\epsilon,\{f_{{\rm L}+}\}\bigr)
   + P_{{\rm L}+} (\mib{r},\epsilon) = 0 ,
               \\
        \label{eq:fL-}
  &  D \cdot \nabla \bigl(  N_{1}^{2}(\mib{r},\epsilon)
                          - R_{2}^{2}(\mib{r},\epsilon) \bigr)
       \nabla f_{{\rm L}-}(\mib{r},\epsilon)
   + I_{\rm L}\bigl(\mib{r},\epsilon,\{f_{{\rm L}-}\}\bigr)
   + P_{{\rm L}-} (\mib{r},\epsilon) = 0 ,
               \\
        \label{eq:fT+}
  &  D \cdot \nabla \bigl(  N_{1}^{2}(\mib{r},\epsilon)
                          + N_{2}^{2}(\mib{r},\epsilon) \bigr)
       \nabla f_{{\rm T}+}(\mib{r},\epsilon)
   - \frac{1}{\tau_{\rm conv}(\epsilon)} f_{{\rm T}+}(\mib{r},\epsilon)
   + I_{\rm T}\bigl(\mib{r},\epsilon,\{f_{{\rm T}+}\}\bigr)
   + P_{{\rm T}+} (\mib{r},\epsilon) = 0 ,
               \\
        \label{eq:fT-}
  &  D \cdot \nabla \bigl(  N_{1}^{2}(\mib{r},\epsilon)
                          + N_{2}^{2}(\mib{r},\epsilon) \bigr)
       \nabla f_{{\rm T}-}(\mib{r},\epsilon)
   - \frac{1}{\tau_{\rm conv}(\epsilon)} f_{{\rm T}-}(\mib{r},\epsilon)
                 \nonumber \\
 & \hspace{10mm}
   - \frac{1}{\tau_{\rm sf}} \bigl(  N_{1}^{2}(\mib{r},\epsilon)
                                   + N_{2}^{2}(\mib{r},\epsilon) \bigr)
                             f_{{\rm T}-}(\mib{r},\epsilon)
   + I_{\rm T}\bigl(\mib{r},\epsilon,\{f_{{\rm T}-}\}\bigr)
   + P_{{\rm T}-} (\mib{r},\epsilon) = 0 ,
\end{align}
where $\tau_{\rm sf}^{-1} = (4/3)\tau_{\rm so}^{-1}$ and
\begin{align}
   P_{{\rm L}\pm}(\mib{r},\epsilon)
& = \frac{1}{2}\left(  P_{\rm L}(\mib{r},\epsilon)
                       \pm P_{\rm L}(\mib{r},-\epsilon) \right) ,
         \\
   P_{{\rm T}\pm}(\mib{r},\epsilon)
& = \frac{1}{2}\left(  P_{\rm T}(\mib{r},\epsilon)
                       \pm P_{\rm T}(\mib{r},-\epsilon) \right) .
\end{align}
Using the expression of the tunnel resistance,
\begin{align} 
  R_{t}^{-1}
  = 2\pi e^{2}N_{\rm S}(0)\bigl(N_{{\rm F}\uparrow}(0)+N_{{\rm F}\downarrow}(0)\bigl)
    |\mathcal{T}|^{2} ,
\end{align}
we can rewrite the injection terms as
\begin{align}
       \label{eq:PL+}
   P_{{\rm L}+}(\mib{r},\epsilon)
& = \frac{P_{\rm s}N_{1}(\mib{r},\epsilon)}{8e^{2}N_{\rm S}(0)R_{t}}
    \delta(\mib{r}-\mib{r}_{0})
                    \Bigl(  \tanh\bigl(\frac{\epsilon+eV}{2T}\bigr)
                          - \tanh\bigl(\frac{\epsilon-eV}{2T}\bigr) \Bigr) ,
        \\
      \label{eq:PL-}
   P_{{\rm L}-}(\mib{r},\epsilon)
& = \frac{N_{1}(\mib{r},\epsilon)}{8e^{2}N_{\rm S}(0)R_{t}}
    \delta(\mib{r}-\mib{r}_{0})
                    \Bigl(2 \tanh\bigl(\frac{\epsilon}{2T}\bigr)
                          - \tanh\bigl(\frac{\epsilon+eV}{2T}\bigr)
                          - \tanh\bigl(\frac{\epsilon-eV}{2T}\bigr) \Bigr) ,
        \\
      \label{eq:PT+}
   P_{{\rm T}+}(\mib{r},\epsilon)
& = \frac{N_{1}(\mib{r},\epsilon)}{8e^{2}N_{\rm S}(0)R_{t}}
    \delta(\mib{r}-\mib{r}_{0})
                    \Bigl(  \tanh\bigl(\frac{\epsilon+eV}{2T}\bigr)
                          - \tanh\bigl(\frac{\epsilon-eV}{2T}\bigr) \Bigr) ,
        \\
      \label{eq:PT-}
   P_{{\rm T}-}(\mib{r},\epsilon)
& = \frac{P_{\rm s}N_{1}(\mib{r},\epsilon)}{8e^{2}N_{\rm S}(0)R_{t}}
    \delta(\mib{r}-\mib{r}_{0})
                    \Bigl(2 \tanh\bigl(\frac{\epsilon}{2T}\bigr)
                          - \tanh\bigl(\frac{\epsilon+eV}{2T}\bigr)
                          - \tanh\bigl(\frac{\epsilon-eV}{2T}\bigr) \Bigr) ,
\end{align}
where the spin polarization $P_{\rm s}$ is defined by
\begin{align}
 P_{\rm s} = \frac{N_{{\rm F}\uparrow}(0)-N_{{\rm F}\downarrow}(0)}
                  {N_{{\rm F}\uparrow}(0)+N_{{\rm F}\downarrow}(0)} .
\end{align}

Equations~(\ref{eq:fL+})-(\ref{eq:fT-})
and eqs.~(\ref{eq:PL+})-(\ref{eq:PT-}) are the central result of this paper.
We here clarify the meaning of each distribution function.
The distribution functions $f_{\sigma}(\mib{r},\epsilon)$ are expressed as
\begin{align}
         \label{eq:f-up2}
     f_{\uparrow}(\mib{r},\epsilon)
 & = f_{\rm FD}(\epsilon,\mu)
         + f_{{\rm L}+}(\mib{r},\epsilon) + f_{{\rm L}-}(\mib{r},\epsilon)
         + f_{{\rm T}+}(\mib{r},\epsilon) + f_{{\rm T}-}(\mib{r},\epsilon) ,
            \\
         \label{eq:f-down2}
     f_{\downarrow}(\mib{r},\epsilon)
 & = f_{\rm FD}(\epsilon,\mu)
         - f_{{\rm L}+}(\mib{r},\epsilon) + f_{{\rm L}-}(\mib{r},\epsilon)
         + f_{{\rm T}+}(\mib{r},\epsilon) - f_{{\rm T}-}(\mib{r},\epsilon) .
\end{align}
Let $S(\mib{r})$ and $Q(\mib{r})$ be the spin and charge imbalances,
respectively.
Noting that $N_{1}(\mib{r},\epsilon)$ is the normalized local density of
states in the superconductor, we can express
\begin{align}
   S(\mib{r})
 & = N_{\rm S}(0)\int_{-\infty}^{\infty}{\rm d}\epsilon \hspace{1mm}
     N_{1}(\mib{r},\epsilon)
     \left( f_{\uparrow}(\mib{r},\epsilon) - f_{\downarrow}(\mib{r},\epsilon)
     \right) ,
            \\
   Q(\mib{r})
 & = N_{\rm S}(0)\int_{-\infty}^{\infty}{\rm d}\epsilon \hspace{1mm}
     N_{1}(\mib{r},\epsilon)
     \left( f_{\uparrow}(\mib{r},\epsilon) + f_{\downarrow}(\mib{r},\epsilon)
             - 2 f_{\rm FD}(\epsilon,\mu) \right) .
\end{align}
By using eqs.~(\ref{eq:f-up2}) and (\ref{eq:f-down2}),
we obtain
\begin{align}
     S(\mib{r})
 & = 4 N_{\rm S}(0)\int_{0}^{\infty}{\rm d}\epsilon \hspace{1mm}
     N_{1}(\mib{r},\epsilon) f_{{\rm L}+}(\mib{r},\epsilon) ,
          \\
     Q(\mib{r})
 & = 4 N_{\rm S}(0)\int_{0}^{\infty}{\rm d}\epsilon \hspace{1mm}
    N_{1}(\mib{r},\epsilon) f_{{\rm T}+}(\mib{r},\epsilon) .
\end{align}
Thus, $f_{{\rm L}+}$ and $f_{{\rm T}+}$ describe the spin and charge
imbalances, respectively.
Other two distribution functions are related to quasiparticle energies.
Let $E_{Q}$ and $E_{S}$ be the excess quasiparticle energy and
the energy imbalance between up- and down-spin quasiparticles,
respectively.
They are given by
\begin{align}
   E_{Q}(\mib{r})
 & = N_{\rm S}(0)\int_{-\infty}^{\infty}{\rm d}\epsilon \hspace{1mm}
     N_{1}(\mib{r},\epsilon) \epsilon
     \left( f_{\uparrow}(\mib{r},\epsilon) + f_{\downarrow}(\mib{r},\epsilon)
              - 2 f_{\rm FD}(\epsilon,\mu)
     \right) ,
            \\
   E_{S}(\mib{r})
 & = N_{\rm S}(0)\int_{-\infty}^{\infty}{\rm d}\epsilon \hspace{1mm}
     N_{1}(\mib{r},\epsilon) \epsilon
     \left( f_{\uparrow}(\mib{r},\epsilon) - f_{\downarrow}(\mib{r},\epsilon)
     \right) .
\end{align}
By using eqs.~(\ref{eq:f-up2}) and (\ref{eq:f-down2}),
we obtain
\begin{align}
     E_{Q}(\mib{r})
 & = 4 N_{\rm S}(0)\int_{0}^{\infty}{\rm d}\epsilon \hspace{1mm}
     N_{1}(\mib{r},\epsilon) \epsilon f_{{\rm L}-}(\mib{r},\epsilon) ,
          \\
     E_{S}(\mib{r})
 & = 4 N_{\rm S}(0)\int_{0}^{\infty}{\rm d}\epsilon \hspace{1mm}
     N_{1}(\mib{r},\epsilon) \epsilon f_{{\rm T}-}(\mib{r},\epsilon) .
\end{align}
Thus, $f_{{\rm L}-}$ and $f_{{\rm T}-}$ describe the excess quasiparticle
energy and the energy imbalance between up- and down-spin
quasiparticles, respectively.
The two distribution functions $f_{{\rm L}+}$ and $f_{{\rm T}-}$
characterize the spin imbalance, while other two distribution functions
$f_{{\rm L}-}$ and $f_{{\rm T}+}$ characterize the charge imbalance.
The former two have not been discussed in literatures.

We approximately obtain the spectral functions.
The presence of spin-orbit scattering does not result in any changes
of the spectral properties of superconductors.
According to the approximation adopted in eq.~(\ref{eq:G-NR}), we obtain
\begin{align}
 \Delta(\mib{r})
   = N_{\rm S}(0)g^{2}\int_{0}^{\infty} {\rm d}\epsilon \hspace{1mm}
     R_{2}(\mib{r},\epsilon)
     \left( \tanh\bigl(\frac{\epsilon}{2T}\bigr)
              - 2 f_{{\rm L}-}(\mib{r},\epsilon)
     \right) .
\end{align}
This indicates that the suppression of the pair potential due to
a current injection is governed by $f_{{\rm L}-}$.
In contrast, the variation of the phase $\chi$ is related to $f_{{\rm T}+}$
although this point is out of our scope.
We assume that $\Delta$ and $\chi$ spatially vary much slower
than $f_{{\rm L}\pm}$ and $f_{{\rm T}\pm}$.
Thus, we approximate that $\nabla \hat{G}^{R,A} = 0$.
Furthermore, we neglect the phonon self-energy in deriving $\hat{G}^{R,A}$.
The local density of states vanishes for $|\epsilon| < \Delta(\mib{r})$
in this case, so we consider the energy region of
$|\epsilon| \ge \Delta(\mib{r})$ in the following.
After these simplifications, we obtain
\begin{align}
      \label{eq:N1}
      N_{1}(\mib{r},\epsilon)
 &  = \frac{|\epsilon|}{\sqrt{\epsilon^{2}-\Delta^{2}(\mib{r})}} ,
       \\
      \label{eq:R2}
      R_{2}(\mib{r},\epsilon)
 &  = \frac{{\rm sign}(\epsilon) \Delta(\mib{r})}
           {\sqrt{\epsilon^{2}-\Delta^{2}(\mib{r})}} ,
\end{align}
and $N_{2}(\mib{r},\epsilon)=0$ for $|\epsilon| \ge \Delta(\mib{r})$.

In the presence of the spin and/or charge imbalances, the distribution
functions $f_{\sigma}(\mib{r},\epsilon)$ deviate from the equilibrium ones.
To characterize their deviations,
we introduce the spin-dependent chemical potential
$\mu_{\sigma}(\mib{r}) = \mu + \delta\mu_{\sigma}(\mib{r})$ for quasiparticles.
To define $\delta\mu_{\sigma}(\mib{r})$, we assume that a fictitious electrode
is weakly coupled to a superconductor at $\mib{r}$
through a point-like tunnel junction.
In terms of a bias voltage $V_{\rm fic}$, the spin-dependent tunneling current
$I_{\sigma}(\mib{r},V_{\rm fic})$ is given by
\begin{align}
     I_{\uparrow,\downarrow}(\mib{r},V_{\rm fic})
 &   \propto \int_{0}^{\infty} {\rm d}\epsilon \hspace{1mm}
     N_{1}(\mib{r},\epsilon)
     \Bigl(  \tanh\bigl(\frac{\epsilon+eV_{\rm fic}}{2T} \bigr)
           - \tanh\bigl(\frac{\epsilon-eV_{\rm fic}}{2T} \bigr)
               \nonumber \\
 & \hspace{40mm}
         - 4 \left( \pm f_{{\rm L}+}(\mib{r},\epsilon)
                      - f_{{\rm T}+}(\mib{r},\epsilon) \right)
   \Bigr) .
\end{align}
The tunneling current $I_{\sigma}(\mib{r},V_{\rm fic})$ vanishes
if $eV_{\rm fic}$ is equal to $\delta\mu_{\sigma}(\mib{r})$.
Thus, $\delta\mu_{\sigma}(\mib{r})$ satisfies the following equation
\begin{align}
      \label{eq:def-qcp}
 &  \int_{0}^{\infty} {\rm d}\epsilon \hspace{1mm} N_{1}(\mib{r},\epsilon)
    \Bigl(  \tanh\bigl(\frac{\epsilon+\delta\mu_{\uparrow,\downarrow}(\mib{r})}
                            {2T}
                 \bigr)
          - \tanh\bigl(\frac{\epsilon-\delta\mu_{\uparrow,\downarrow}(\mib{r})}
                            {2T}
                \bigr)
               \nonumber \\
 & \hspace{50mm}
          - 4 \left( \pm f_{{\rm L}+}(\mib{r},\epsilon)
                       - f_{{\rm T}+}(\mib{r},\epsilon) \right)
   \Bigr) = 0 .
\end{align}
We use the above equation as the definition of $\delta\mu_{\sigma}(\mib{r})$.
We observe that
$\delta\mu_{\uparrow}(\mib{r}) = \delta\mu_{\downarrow}(\mib{r})$
if the spin imbalance is absent (i.e., $f_{{\rm L}+}(\mib{r},\epsilon) = 0$),
while $\delta\mu_{\uparrow}(\mib{r}) = - \delta\mu_{\downarrow}(\mib{r})$
in the absence of the charge imbalance
(i.e., $f_{{\rm T}+}(\mib{r},\epsilon) = 0$).

\section{Spin-Imbalance Relaxation}

In this section, we study the behavior of spin-imbalance relaxation
based on the Boltzmann equation for $f_{{\rm L}+}$.
Let us consider a thin superconducting wire coupled with a ferromagnetic
electrode through a tunnel junction.
We assume that $f_{{\rm L}-}$ is very small everywhere
in the superconductor, and set $\Delta(\mib{r}) = \Delta$.
We neglect the charge imbalance for simplicity,
and focus on a shift of the spin-dependent chemical potential.
If the cross-sectional area $A$ of the wire is small enough, we are allowed
to consider a one-dimensional problem of $f_{{\rm L}+}(x,\epsilon)$.
As noted just below eq.~(\ref{eq:def-qcp}), we observe that
$\delta\mu (x) \equiv \delta\mu_{\uparrow}(x) = - \delta\mu_{\uparrow}(x)$
in this case.
The Boltzmann equation for $f_{{\rm L}+}$ is reduced to
\begin{align}
        \label{eq:fL+1D}
  D \partial_{x}^{2} f_{{\rm L}+}(x,\epsilon)
   - \frac{1}{\tau_{\rm sf}} f_{{\rm L}+}(x,\epsilon)
   + I_{\rm L}\bigl(x,\epsilon,\{f_{{\rm L}+}\}\bigr)
   + \tilde{P}_{{\rm L}+} (x,\epsilon) = 0 ,
\end{align}
where
\begin{align}
        \label{eq:IL+sym} 
  I_{\rm L}\bigl(x,\epsilon,\{f\}\bigr)
 & = - 2 \int {\rm d} \epsilon' \sigma_{\rm ph}(\epsilon,\epsilon')
         \left(  N_{1}(\epsilon)N_{1}(\epsilon')
               - R_{2}(\epsilon)R_{2}(\epsilon') \right)
             \nonumber \\
 & \hspace{5mm} \times
     \frac{\cosh^{2}\bigl(\frac{\epsilon}{2T}\bigr)f(x,\epsilon)
           - \cosh^{2}\bigl(\frac{\epsilon'}{2T}\bigr)
                                               f(x,\epsilon')}
          {\sinh\bigl(\frac{\epsilon'-\epsilon}{2T}\bigr)
           \cosh\bigl(\frac{\epsilon}{2T}\bigr)
           \cosh\bigl(\frac{\epsilon'}{2T}\bigr)} ,
             \\
   \tilde{P}_{{\rm L}+}(x,\epsilon)
 & = \frac{P_{\rm s}N_{1}(\epsilon)}{8e^{2}N_{\rm S}(0)R_{t}A}
     \delta(x-x_{0})\Bigl(  \tanh\bigl(\frac{\epsilon+eV}{2T}\bigr)
                          - \tanh\bigl(\frac{\epsilon-eV}{2T}\bigr) \Bigr) .
\end{align}
We define the energy relaxation time $\tau_{\rm ene}(\epsilon)$ as
\begin{align}
  \frac{1}{\tau_{\rm ene}(\epsilon)}
  =  2 \int {\rm d} \epsilon' \sigma_{\rm ph}(\epsilon,\epsilon')
         \left(  N_{1}(\epsilon)N_{1}(\epsilon')
               - R_{2}(\epsilon)R_{2}(\epsilon') \right)
     \frac{\cosh\bigl(\frac{\epsilon}{2T}\bigr)}
          {\sinh\bigl(\frac{\epsilon'-\epsilon}{2T}\bigr)
           \cosh\bigl(\frac{\epsilon'}{2T}\bigr)} ,
\end{align}
in terms of which the first term in the collision integral $I_{\rm L}$
is rewritten as $-f_{{\rm L}+}(\mib{r},\epsilon)/\tau_{\rm ene}(\epsilon)$.
Note that the energy of an injected quasiparticle is within
$\Delta \le |\epsilon| \lesssim eV + T$.
The behavior of the spin-imbalance relaxation depends on whether
$\tau_{\rm sf}$ is longer or shorter than $\tau_{\rm ene}(\epsilon)$
in this energy range.

We first consider the high-temperature regime in which $eV \ll T$ and
$\tau_{\rm ene}(\epsilon)$ is much shorter than $\tau_{\rm sf}$ for
$\Delta \le |\epsilon| \lesssim T$.
In this case, the energy dependence of $f_{{\rm L}+}$ is mainly determined by
the collision-integral term.
Except near the injection point (i.e., $x = x_{0}$),
we can approximate
$f_{{\rm L}+}(x,\epsilon) \propto 1/\cosh^{2}(\epsilon/(2T))$.
Since eq.~(\ref{eq:def-qcp}) is simplified to
\begin{align}
   \int_{0}^{\infty} {\rm d}\epsilon \hspace{1mm} N_{1}(\epsilon)
   \Bigl(  \frac{\delta\mu (x)}{T \cosh^{2}\bigl(\frac{\epsilon}{2T}\bigr)}
         - 4 f_{{\rm L}+}(x,\epsilon) \Bigr) = 0 ,
\end{align}
we observe that
\begin{align}
  f_{{\rm L}+}(x,\epsilon)
  = \frac{\delta\mu (x)}{4T\cosh^{2}\bigl(\frac{\epsilon}{2T}\bigr)} .
\end{align}
We determine $\delta\mu(x)$ based on eq.~(\ref{eq:fL+1D}).
If we approximately set $N_{1}(\epsilon) = 1$ in
$\tilde{P}_{{\rm L}+}(x,\epsilon)$, we obtain
\begin{align}
  \delta\mu (x) = \delta\mu_{0} \hspace{1mm}
                  {\rm e}^{-\frac{|x-x_{0}|}{\lambda_{\rm s}}} ,
\end{align}
where $\lambda_{\rm s}=\sqrt{D\tau_{\rm sf}}$ is the spin-diffusion length and
\begin{align}
  \delta\mu_{0}
   = \frac{2\lambda_{\rm s}P_{\rm s}eV}{8e^{2}N_{\rm S}(0)R_{t}AD} .
\end{align}
The approximation $N_{1}(\epsilon) \to 1$ results in an under-estimation of
$\delta\mu_{0}$.
It should be noted that the distribution function
$f_{\uparrow,\downarrow}(x,\epsilon) \equiv f_{\rm FD}(\epsilon,\mu)
\pm f_{{\rm L}+}(\epsilon,\mu)$ is expressed by the Fermi-Dirac distribution
function with a shifted chemical potential as
$f_{\uparrow,\downarrow}(x,\epsilon)
\approx f_{\rm FD}(\epsilon,\mu \pm \delta\mu (x))$ in this case.

Next, we consider the low-temperature regime in which
$\tau_{\rm ene}(\epsilon)$ is much longer than $\tau_{\rm sf}$ for
$\Delta \le |\epsilon| \lesssim eV$.
Thus, we can neglect the collision-integral term in eq.~(\ref{eq:fL+1D}).
Solving eq.~(\ref{eq:fL+1D}), we obtain
\begin{align}
      \label{eq:fL+low}
  f_{{\rm L}+}(x,\epsilon)
   = f_{0}(\epsilon) \hspace{1mm}
     {\rm e}^{-\frac{|x-x_{0}|}{\lambda_{\rm s}}}
\end{align}
with
\begin{align}
  f_{0}(\epsilon)
  = \frac{\lambda_{\rm s}P_{\rm s}}{16e^{2}N_{\rm S}(0)R_{t}AD}
    N_{1}(\epsilon)
    \left( \tanh\bigl(\frac{\epsilon+eV}{2T} \bigr)
           - \tanh\bigl(\frac{\epsilon-eV}{2T} \bigr) \right) .
\end{align}
In this case, we cannot express the distribution function
$f_{\sigma}(x,\epsilon)$ by the Fermi-Dirac distribution
with a shifted chemical potential.
Substituting eq.~(\ref{eq:fL+low}) into eq.~(\ref{eq:def-qcp}),
we obtain
\begin{align}
       \label{eq:delta-mu-lt}
   \int_{0}^{\infty} {\rm d}\epsilon \hspace{1mm} N_{1}(\epsilon)
   \Bigl(  \tanh\bigl(\frac{\epsilon+\delta\mu(x)}{2T}
                \bigr)
         - \tanh\bigl(\frac{\epsilon-\delta\mu(x)}{2T}
                \bigr)
         - 4 f_{0}(\epsilon) \hspace{1mm}
             {\rm e}^{-\frac{|x-x_{0}|}{\lambda_{\rm s}}} \Bigr) = 0 .
\end{align}
We numerically solve eq.~(\ref{eq:delta-mu-lt}) and obtain
$\delta\mu(x)/\Delta$ as a function of the normalized distance
$y \equiv |x-x_{0}|/\lambda{\rm s}$ from the injection point $x_{0}$
at $eV/\Delta = 1.2$ and $1.6$ for $T/\Delta = 0.04$, $0.08$ and $0.16$.
The following parameters are adopted:
$R_{t} = 2 \hspace{1mm} {\rm k}\Omega$,
$N_{\rm S}(0) = 1.2 \times 10^{22} \hspace{1mm} {\rm eV}^{-1}{\rm cm}^{-3}$,
$D = 5.3 \times 10^{9} \hspace{1mm} \mu {\rm m}^{2}/{\rm s}$,
$\tau_{\rm sf} = 200 \hspace{1mm} {\rm ps}$ and
$A = 50 \times 250 \hspace {1mm} {\rm nm}^{2}$.
The values of $D$ and $\tau_{\rm sf}$ result
in $\lambda_{\rm s} = 1.03 \hspace{1mm} \mu {\rm m}$.
\begin{figure}[btp]
\begin{center}
\includegraphics[height=8cm]{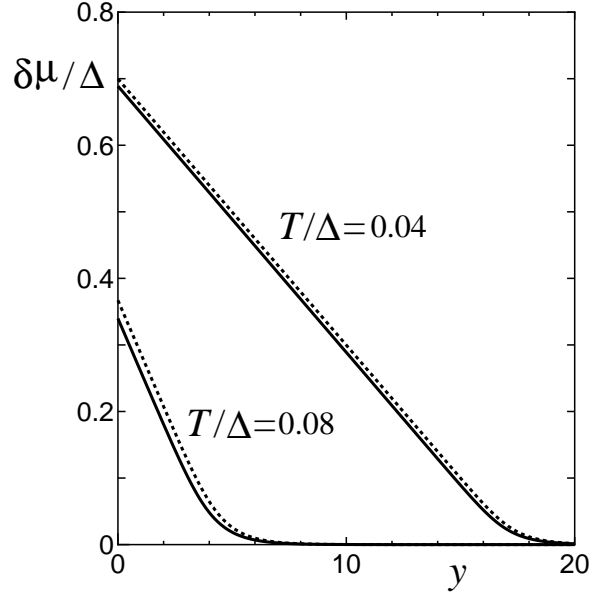}
\end{center}
\caption{$\delta\mu / \Delta$ for $T/\Delta = 0.04$ and $0.08$
as a function of the normalized distance $y \equiv |x-x_{0}|/\lambda_{\rm s}$
from the injection point.
The solid and dotted lines correspond to
$eV/\Delta = 1.2$ and $1.6$, respectively.
}
\end{figure}
\begin{figure}[btp]
\begin{center}
\includegraphics[height=8cm]{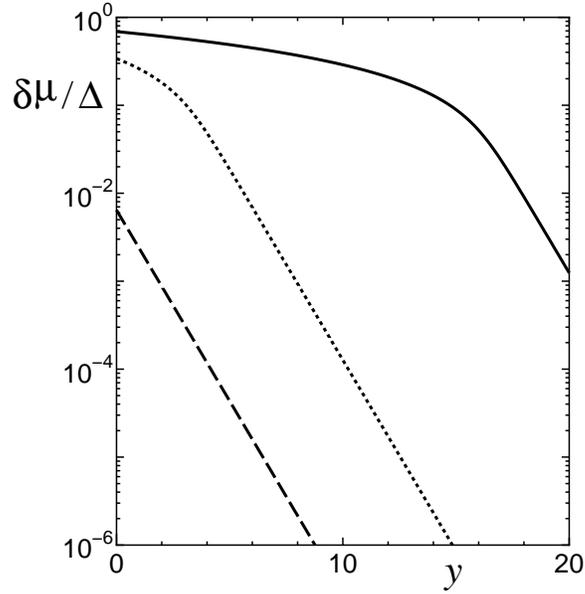}
\end{center}
\caption{$\delta\mu / \Delta$ at $eV/\Delta = 1.2$ as a function of
the normalized distance $y \equiv |x-x_{0}|/\lambda_{\rm s}$
from the injection point.
The solid, dotted and dashed lines correspond to
$T/\Delta = 0.04$, $0.08$ and $0.16$, respectively.
}
\end{figure}
Figure~1 shows that the bias-voltage dependence of $\delta \mu$ is very weak,
in contrast to the high-temperature regime
where $\delta \mu \propto eV$.
This clearly indicates a nonlinear nature of eq.~(\ref{eq:delta-mu-lt}).
From Fig.~2, we observe that the decay of $\delta\mu(x)$ can be
fitted by the exponential law ${\rm e}^{- |x-x_{0}|/\lambda_{\rm s}}$
in the case of $T/\Delta = 0.16$.
However, except for this case, a deviation from
the exponential law appears near the injection point (i.e., $y = 0$).
We also observe that the asymptotic behavior of $\delta\mu(x)$
far from the injection point is still governed by the exponential law.
It should be noted here that the anomalous slow decay of $\delta\mu$ observed
near the injection point is partly attributed to the divergence of
$N_{1}(\epsilon)$ at the gap edge.
Since the divergence is smeared by a gap anisotropy,
the anomalous behavior may be weakened in actual cases.

Yamashita \textit{et al.}~\cite{rf:yamashita} have studied the spin-imbalance
relaxation in superconductors by assuming that the quasiparticle distribution
function can be expressed in the form of the Fermi-Dirac distribution function
with a shifted chemical potential $\mu_{\sigma}(x) = \mu \pm \delta\mu (x)$,
and concluded that $\delta\mu_{\sigma}(x)$ decays exponentially
on the length scale of $\lambda_{\rm s}$.
Our argument indicates that the assumption employed in
ref.~\citen{rf:yamashita} can be justified only in the high-temperature regime.
Indeed, near the injection point at low temperatures, we have found
deviations from the exponential law.

We have neglected the charge imbalance in the above argument.
The nonlinear equation, eq.~(\ref{eq:def-qcp}), is reduced to
\begin{align}
     \delta\mu_{\uparrow, \downarrow}(x)
  =   4 T \cosh^{2}\bigl(\frac{\epsilon}{2T}\bigr)
        \bigl(\pm f_{{\rm L}+}(x,\epsilon) + f_{{\rm T}+}(x,\epsilon) \bigr)
\end{align}
in the high-temperature regime.
Thus, an additional spin-independent correction is simply added to
the chemical potential if we take the charge imbalance into account.
However, such a simple treatment cannot be applied to
the low-temperature regime because we must solve eq.~(\ref{eq:def-qcp})
in its present form to obtain $\delta\mu_{\sigma}$.
Thus, the spin-imbalance and charge-imbalance corrections to
$\delta\mu_{\sigma}$ are not necessarily additive.

\section{Summary}

We have studied a nonequilibrium distribution of quasiparticles
in the presence of both spin and charge imbalances.
By extending the kinetic equation approach by Schmid and Sch\"{o}n
based on the quasiclassical Green's function method,
we have presented a set of linearized Boltzmann equations
for distribution functions $f_{{\rm L}\pm}(\mib{r},\epsilon)$ and
$f_{{\rm T}\pm}(\mib{r},\epsilon)$ in steady states.
It is shown that $f_{{\rm L}+}$ and $f_{{\rm T}-}$ represent
the spin and charge imbalances, respectively, and $f_{{\rm L}-}$ and
$f_{{\rm T}+}$ represent the excess quasiparticle energy and the energy
imbalance between up- and down-spin quasiparticles, respectively.
It is also shown that the suppression of the pair potential due to
a current injection is governed by $f_{{\rm L}-}$.
These distribution functions are decoupled with each other
in the Boltzmann equations.
This allows us to separately consider the spin and charge imbalances
in steady states.

As an application of the Boltzmann equations,
we have considered the relaxation of spin imbalance in a quasi-one-dimensional
superconducting wire weakly coupled with a ferromagnetic electrode.
The spin imbalance induces a shift $\delta\mu$ ($- \delta \mu$) of
the chemical potential for up-spin (down-spin) quasiparticles.
We have analyzed spatial decay of $\delta \mu$
due to spin-orbit impurity scattering.
We have shown that at high temperatures, $\delta\mu$ decays exponentially
on the length scale of $\lambda_{\rm s}$,
where $\lambda_{\rm s}$ is the spin diffusion length.
However, at low temperatures,
we have observed deviations from the exponential law near the injection point.


\begin{thebibliography}{99}


\bibitem{rf:rieger} T. J. Rieger, D. J. Scalapino and J. E. Mercereau:
Phys. Rev. Lett. {\bf 27} (1971) 1787.

\bibitem{rf:yu1} M. L. Yu and J. E. Mercereau:
Phys. Rev. Lett. {\bf 28} (1972) 1117.

\bibitem{rf:clarke1} J. Clarke:
Phys. Rev. Lett. {\bf 28} (1972) 1363.

\bibitem{rf:tinkham1} M. Tinkham and J. Clarke:
Phys. Rev. Lett. {\bf 28} (1972) 1366.

\bibitem{rf:tinkham2} M. Tinkham:
Phys. Rev. B {\bf 6} (1972) 1747.

\bibitem{rf:clarke2} J. Clarke and J. L. Paterson:
J. Low Tem. Phys. {\bf 15} (1974) 491.

\bibitem{rf:schmid} A. Schmid and G. Sch{\"o}n:
J. Low Tem. Phys. {\bf 20} (1975) 207.

\bibitem{rf:yu2} M. L. Yu and J. E. Mercereau:
Phys. Rev. B {\bf 12} (1975) 4909.

\bibitem{rf:ikebuchi} Y. Ikebuchi and R. Yagi: 
Physica E {\bf 22} (2004) 757.

\bibitem{rf:tedrow} P. M. Tedrow and R. Meservey:
Phys. Rev. B {\bf 7} (1973) 318.

\bibitem{rf:johnson} M. Johson:
Appl. Phys. Lett. {\bf 65} (1994) 1460.

\bibitem{rf:urech} M. Urech, J. Johansson, V. Korenivski and D. B. Haviland:
J. Magn. Magn. Mater. {\bf 272-276} (2004) Suppl. E1469.

\bibitem{rf:shin} Y.-S. Shin, H.-J. Lee and H.-W. Lee:
Phys. Rev. B {\bf 71} (2005) 144513.

\bibitem{rf:zhao} H. L. Zhao and S. Hershfield:
Phys. Rev. B {\bf 52} (1995) 3632.

\bibitem{rf:yamashita} T. Yamashita, S. Takahashi, H. Imamura and
S. Maekawa: Phys. Rev. B {\bf 65} (2002) 172509.

\bibitem{rf:kopnin} N. Kopnin:
\textit{Theory of Nonequilibrium Superconductivity}
(Oxford University Press, Oxford, 2001).

\bibitem{rf:eilenberger} G. Eilenberger:
Z. Phys. {\bf 214} (1968) 195.

\bibitem{rf:eliashberg} G. M. Eliashberg:
Sov. Phys. JETP {\bf 34} (1972) 668.

\bibitem{rf:larkin} A. I. Larkin and Yu. N. Ovchinnikov:
Sov. Phys. JETP {\bf 46} (1977) 155.

\bibitem{rf:usadel} K. D. Usadel:
Phys. Rev. Lett. {\bf 25} (1970) 507.

\bibitem{rf:abrikosov} A. A. Abrikosov and L. P. Gorkov:
Sov. Phys. JETP {\bf 15} (1962) 752.

\bibitem{rf:takane} See the discussion in Y. Takane:
J. Phys. Soc. Jpn. {\bf 75} (2006) 023706.

\end{thebibliography}
\end{document}